\documentclass[runningheads]{llncs}
\usepackage{graphicx}
\usepackage{hyperref}

\usepackage{mathptmx}       
\usepackage{helvet}         
\usepackage{courier}        
\usepackage{type1cm}        
\usepackage{amssymb}
\usepackage{algorithm}
\usepackage{algpseudocode}
\usepackage{tikz,amsmath,listings,textcomp}
\input{listings-maple-definition.sty}
\usepackage{xfrac,subfigure}
\usepackage{makeidx}         
\usepackage{graphicx}        
\usepackage{multicol}        
\usepackage[bottom]{footmisc}

\makeindex             
\newcommand{\Maple}{\textsc{Maple}}
\newcommand{\Blend}{\lstinline!Blend!}

\renewcommand{\emph}[1]{\textsl{#1}}
\usepackage{bm}

\begin{document}
\title{Blends in \Maple\thanks{This work partially funded by NSERC.}}
%
%
\author{Robert M.~Corless\inst{1,2}\orcidID{0000-0003-0515-1572} \and
Erik J.~Postma\inst{3}\orcidID{0000-0003-0059-4163}
}
\authorrunning{R.M.~Corless \and E.~Postma}
%
\institute{School of Mathematical and Statistical Sciences, Western Universty\and
David R.~Cheriton School of Computer Science, Waterloo University
\email{rcorless@uwo.ca}
\and
Maplesoft, Waterloo, Ontario
\email{epostma@maplesoft.com}
}
\maketitle              
\begin{abstract}
A \emph{blend} of two Taylor series for the same smooth real- or complex-valued function of a single variable can be useful for approximation of said function. We use an explicit formula for a two-point Hermite interpolational polynomial to construct such blends.  We show a robust \Maple\ implementation that can stably and efficiently evaluate blends using linear-cost Horner form, evaluate their derivatives to arbitrary order at the same time, or integrate a blend exactly.  The implementation is suited for use with \lstinline!evalhf!.  We provide a top-level user interface and efficient module exports for programmatic use.
\textbf{This work was presented in video form, with our software provided, at the Maple Conference 2020.} \href{https://www.maplesoft.com/mapleconference/}{www.maplesoft.com/mapleconference}
\end{abstract}
\section{Introduction}
Taylor series are one of the basic tools of analysis and of computation for functions of a single variable.  However, outside of specialist circles it is not widely appreciated that two Taylor series can be rapidly and stably combined to give what is usually a much better approximation than either one alone.  In this paper we only consider blending Taylor series at two points, say $z=a$ and $z=b$.  We convert to the unit interval by introducing a new variable~$s$ with $z = a + s(b-a)$.  Most examples in this paper will just use~$s$, but it is a straightforward matter to adjust back to the original variables, and we will give examples of how to do so.
\subsection{The basic formula}
Consider the following formula, known already to Hermite, which states that the grade $m+n+1$ polynomial
\begin{align}
    H_{m,n}(s) = &\sum _{j=0}^{m} \left[ \sum _{k=0}^{m-j}{n+k\choose k}{s}^{k+j}
 \left( 1-s \right) ^{n+1}\right]p_{{j}}  \nonumber\\
 +& \sum _{j=0}^{n}
\left[\sum _{k=0}^{n-j}{m+k\choose k}{s}^{m+1}
 \left( 1-s \right) ^{k+j}\right] \left( -1 \right) ^{j}q_{{j}}  \label{eq:TPHI}
\end{align}
has a Taylor series matching the given $m+1$ values $p_j = f^{(j)}(0)/j!$ at $s=0$ and another Taylor series matching the given $n+1$ values $q_j = f^{(j)}(1)/j!$ at $s=1$.  Putting this in symbolic terms and using a superscript $(j)$ to mean the $j$th derivative with respect to~$s$, we have
\[
\frac{H^{(j)}_{m,n}(0)}{j!} = p_j\>, \qquad 0 \le j \le m\>, \qquad \mathrm{and} \qquad
\frac{H^{(j)}_{m,n}(1)}{j!} = q_j\>, \qquad 0 \le j \le n\>.
\]
This is a kind of interpolation, indeed a special case of what is called \emph{Hermite} interpolation.
As with Lagrange interpolation, where for instance two points give a grade one polynomial, that is, a line, here $m+n+2$ pieces of information gives a grade $m+n+1$ polynomial.  We will see that this formula can be evaluated in $O(m)+O(n)$ arithmetic operations.

We use the word \textsl{grade} to mean ``degree at most''.  That is, a polynomial of grade (say)~$5$ is of degree at most~$5$, but because here the leading coefficient is not visible, we don't know the exact degree, which could be lower.  Typically, with a blend we will not know the degree unless we compute it.  This use of the word ``grade'' is common in the literature of matrix polynomial eigenvalue problems.

\subsection{Applications}
Our initial motivation was in writing code in Maple to solve the Mathieu differential equation in~\cite{brimacombe2020computation} using a Hermite-Obreschkoff method~\cite{nedialkov1999interval,Nedialkov(2005),Nedialkov(2007)}, which uses Taylor series at either end of each numerical step; this implicit high-order method is especially suited to differential equations (such as the Mathieu equation) for which the Taylor series at any point may be computed quickly.  Blends can also be used for quadrature (numerical evaluation of definite integrals), or for approximation of functions.  We will see examples of that last, in the next section.

Using a companion matrix discussed in~\cite{Lawrence:2012:NSB:2331684.2331706}, we can also find approximate zeros of nonlinear functions from Taylor series data at either end of an interval.  This can be turned into an efficient iterative method of order $1+\sqrt{3} > 2$ with the same cost as Newton's method, by using \emph{reversed} Taylor series: see~\cite{corless2020inverse,corless2020pure}.

There are also interesting pedagogical applications of blends.  Using them can strengthen the notion of convergence in students' minds; this can be done at an elementary calculus level or at a real analysis level.  The companion matrix mentioned earlier can be used in Linear Algebra as a topic in computing eigenvalues.  The derivation of the method is a lovely exercise in contour integration for a course in complex variables.  Of course, it provides a topic in approximation theory and in numerical analysis: a proof that the method is numerically stable will be given elsewhere.  Interestingly, blends generate an infinite number of different quadrature rules, unusual members of which can be used as unique tools of student assessment.

\subsection{Initial Examples}

\begin{figure}
\centering     
\subfigure[$H_{9,9}(s)-f(s)$ $15$ Digits]{\label{fig:gam9915}\includegraphics[width=60mm]{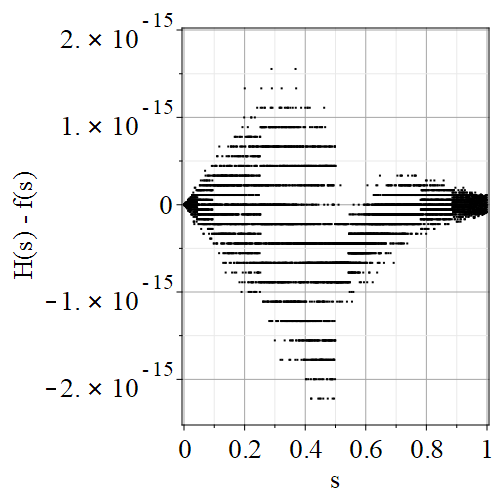}}
\subfigure[$H_{9,9}'(s)-f'(s)$ $15$ Digits]{\label{fig:dgam9915}\includegraphics[width=60mm]{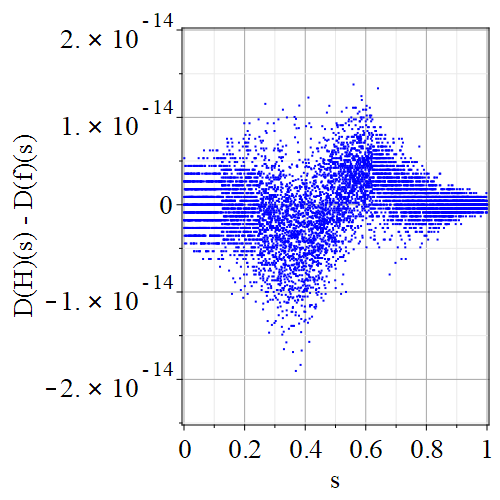}}
\newline
\subfigure[$H_{9,9}(s)-f(s)$ $30$ Digits]{\label{fig:gam9930}\includegraphics[width=60mm]{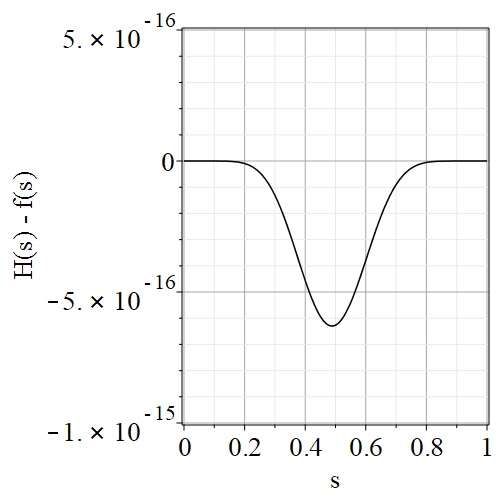}}
\subfigure[$H_{9,9}'(s)-f'(s)$ $30$ Digits]{\label{fig:dgam9930}\includegraphics[width=60mm]{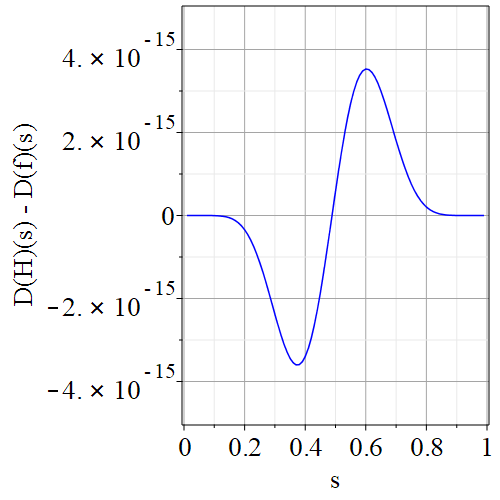}}
\caption{The error in an $(9,9)$ blend for $f(s) = 1/\Gamma(s-3)$.  This grade of blend produces an approximation that is nearly accurate to full double precision; when the correctly rounded reference result is subtracted from the computed result, the truncation error is obscured by rounding errors.  As is shown in the second row of graphs, recomputing these approximations at higher precision gives smoother error curves of about the same size and showing the theoretical $s^{10}(1-s)^{10}$ behaviour. As usual with approximation methods, the accuracy degrades as the derivative order increases. \label{fig:gam99}}
\end{figure}

We show an example in figure~\ref{fig:gam99}.  We take the function $f(s) = 1/\Gamma(s-3)$.  For a reference on the Gamma function, see~\cite{borwein2018gamma}.  This function has known series at $s=0$ and $s=1$:
\[
\frac{1}{\Gamma(s-3)} = -6\,s+ \left( -6\,\gamma+11 \right) {s}^{2}+ \left( {\frac {{\pi}^{2}
}{2}}-3\,{\gamma}^{2}+11\,\gamma-6 \right) {s}^{3}+O \left( {s}^{4}
 \right)
\]
and
\[
\frac{1}{\Gamma(s-3)} =2\, \left( s-1 \right) + \left( 2\,\gamma-3 \right)  \left( s-1
 \right) ^{2}+ \left( -{\frac {{\pi}^{2}}{6}}+{\gamma}^{2}-3\,\gamma+1
 \right)  \left( s-1 \right) ^{3}+O \left(  \left( s-1 \right) ^{4}
 \right)
\]
as computed by \Maple's \lstinline@series@ command\footnote{In fact, \Maple\ can compute the series for $1/\Gamma(z)$ at $z=-n$ where $n$ is a symbol, assumed to be a nonnegative integer.  This example will be discussed further in a separate paper.}.  The series coefficients get complicated as the degree increases, so we suppress printing them.  We compute them up to degrees $m=9$ and $n=9$ and make a blend for this function.  This gives a grade $9+9+1 = 19$ approximation (and indeed the blend turns out to be actually degree $19$; the lead coefficient does not, in fact, cancel).  In the figure, we plot the error $H_{9,9}(s)-f(s)$ and the derivative error $H_{9,9}'(s)-f'(s)$, first in the top row computing the blend in $15$ Digits (which takes a third of a second on a 2018 Surface Pro to compute the blend and three of its derivatives at $2021$ points, so $8084$ values) and then comparing against \Maple's built-in evaluator (computed at higher precision, in fact 60 Digits because of the apparent end-point singularity, and then rounded correctly to $15$ digits).  In the second row we compare the blend computed at $30$ Digits, which takes $3.14$ seconds, about ten times longer than the $15$ Digit computation.  We see in the second row of the figure that the \emph{truncation error}---that is, the error in approximation by taking a degree $19$ polynomial---is smaller than $6\cdot 10^{-16}$; \Maple's hardware floats use IEEE double precision with a unit roundoff of $2^{-53} \approx 10^{-16}$.  We therefore expect rounding error to dominate if we do computation in only $15$ Digits, and that is indeed what we see in the top row---and moreover we see that the rounding error is not apparently amplified very much, if at all: the errors plotted are all modest multiples of the unit roundoff.  The unit roundoff itself can be seen in the apparent horizontal lines, in fact.  This will be indicative of the general behaviour of a blend: when carefully implemented, rounding errors do not affect it much.  Since, as we will see, balanced blends are quite well-conditioned, this will result in usually accurate answers.

To compare with Taylor series and other methods of approximating this particular $f(s)$, an equivalent cost Taylor series would be degree $19$.  The Taylor series of degree $19$ at $s=0$ has an error at $s=1^-$ of about $3.5\cdot 10^{-7}$, many orders of magnitude greater than the error in the blend; the series at $s=1$ has a similar-sized error at $s=0^+$.  This is well-known: Taylor series are really good at their point of expansion, but will be bad at the other end of the interval.  On the other hand, the ``best'' polynomial approximation to this function, best in the minimax sense and found by the Remez algorithm, is of course better than the blend we produce here.  Similarly a Chebyshev approximation to this function, as would be produced by Chebfun~\cite{Battles(2004)}, is also better: either cheaper for the same accuracy, or more accurate for the same effort.  And then there is the new AAA algorithm, which is better still~\cite{nakatsukasa2018aaa}, which we do not pursue further here.  But where does a blend fit in on this scale of best-to-Taylor?  The Chebyshev approximation accurate to $6\cdot 10^{-16}$ (as computed by \lstinline!numapprox[chebyshev]! is of degree $16$, not $19$, so it is about $20$\% cheaper to evaluate\footnote{This is harder to judge than we are saying, here.  Optimal evaluation of Chebyshev polynomials via preprocessing is not usually done; the Clenshaw algorithm is backward stable (see e.g.~\cite{corless2013graduate}) and usually used because it is $O(n)$ in cost.  Similarly, evaluation of a blend is $O(m+n)$.  So this figure of $20$\% is likely not very true, but rather merely indicative.}.  For the rational Remez best approximation, by \lstinline@numapprox[minimax]@ the degree $[8,7]$ gets an error $3\cdot 10^{-16}$ and is cheaper yet to evaluate.
Conversely, when using a single Taylor series, experiments at high precision show that
we need to use degree $29$ to get an error strictly less
than the error of the $(9,9)$ blend everywhere in the interval, and a degree $28$ Taylor series is strictly worse.
%
%
%
%
%
%
%
Therefore both the best approximation and the Chebyshev series are better than a blend---but in this case not by that much, while a blend beats a single Taylor series by a considerable margin.  There are other examples where Chebyshev series beat blends by a similarly large margin, but because blends are relatively simple to compute and to understand, being ``sometimes in the ballpark'' of the best kinds of approximation is likely good enough to make these objects interesting.  We are especially interested in situations where Taylor series at either end of an interval are known or very cheap to compute, e.g. for so-called \emph{holonomic} or \emph{D-finite} functions~\cite{van2001fast,mezzarobba2010numgfun,mezzarobba2012note}.  Note that even there we find that sometimes Chebyshev series are worth the extra effort~\cite{benoit2017rigorous}.
\section{Truncation error and rounding error}
The error in Hermite interpolation is known, see for instance~\cite{Kansy(1973)}.  Here, the general real results simplify to
\begin{equation}
f(s) - H_{m,n}(s) = \frac{f^{(m+n+2)}(\theta)}{(m+n+2)!} s^{m+1}(s-1)^{n+1} \label{eq:errorblend}
\end{equation}
for some $\theta = \theta(s)$ between $0$ and $1$.  This is quite reminiscent of the Lagrange form of the remainder of Taylor series, and indeed it reduces exactly to that if we have an $(m,-1)$ or $(-1,n)$ blend---that is, without using any information from the other point.  We saw in the high-precision graphs in figure~\ref{fig:gam99} that the actual error curve really does flatten out at both ends, when information is known at both ends.

The errors in derivatives have a similar form, and as shown in~\cite{Kansy(1973)} essentially lose only one order of accuracy for each derivative taken.

The numbers $\binom{n+k}{k}$, for $0 \le k \le m$, and identically $\binom{m+k}{k}$, $0 \le k \le n$, which appear in formula~\eqref{eq:TPHI}, grow large rather quickly.  Here is a table of the numbers: to get $\binom{m+k}{k}$, choose the~$m$th row (indexed starting at $0$) and read across from $k=0$ until $k=n$.  To get $\binom{n+k}{k}$, choose the~$n$th column and read down from $k=0$ until $k=m$.  Of course the table is symmetric.
\[
 \left[ \begin {array}{cccccccccc} 1&1&1&1&1&1&1&1&1&\cdots
\\ \noalign{\medskip}1&2&3&4&5&6&7&8&9&\cdots
\\ \noalign{\medskip}1&3&6&10&15
&21&28&36&45&\cdots
\\ \noalign{\medskip}1&4&10&20&35&56&84&120&165&\cdots
\\ \noalign{\medskip}1&5&15&35&70&126&210&330&495&\cdots
\\ \noalign{\medskip}
1&6&21&56&126&252&462&792&1287&\cdots
\\ \noalign{\medskip}1&7&28&84&210&462&
924&1716&3003&\cdots
\\ \noalign{\medskip}1&8&36&120&330&792&1716&3432&6435
\\ \noalign{\medskip}1&9&45&165&495&1287&3003&6435&12870&\cdots\\ \noalign{\medskip}
\vdots &\vdots &\vdots &\vdots &\vdots &\vdots &\vdots &\vdots &\vdots & \ddots
\end {array}
 \right]
\]
The largest entries are on the diagonal, and indeed
\begin{equation}\label{eq:numbergrowth}
  \binom{2m}{m} \sim \frac{4^m}{\sqrt{\pi m}}\left( 1 + O\left(\frac{1}{m}\right)\right)
\end{equation}
as we find out from the \Maple\  command
\begin{lstlisting}
asympt( binomial(2*m,m), m )
\end{lstlisting}
and some simplification.  One worries about the numerical effect of these large numbers for high-degree blends.
These \emph{do} have some bad numerical effects sometimes, such as in the companion matrices of~\cite{Lawrence:2012:NSB:2331684.2331706}, but we will also see by the example of high-degree blends that their influence is not as bad as it might have been feared.  For instance, when $m=n=233$, we have $\binom{2m}{m} = 7\cdot 10^{138}$.  Yet the blends that we have computed of this grade (including those for the Lebesgue function, in the next section) show no numerical difficulties at all.

Indeed rounding error depends strongly on how the blend is actually evaluated.  The ordinary Horner's rule has a standard ``backward error'' result: each evaluation is the \emph{exact} evaluation of a polynomial with only slightly different (in a relative error sense) coefficients. Different $x$ gets different polynomials, of course.  In precise terms, if $p(x) = \sum_{k=0}^{\ell} a_k x^k$ and ``fl'' means the result on floating-point evaluation using Horner's method in IEEE arithmetic with unit roundoff $\mu$ ( in double precision, $\mu = 2^{-53} \approx 10^{-16}$) then, if $2\ell < 1/\mu$,
\begin{equation}\label{eq:Hornerbackward}
 \textrm{fl}\left( p(x) \right) = a_0(1+\theta_1) + \sum_{k=0}^{\ell-1} a_k(1+\theta_{2k+1}) x^k + a_\ell(1+\theta_{2\ell}) x^\ell
\end{equation}
where each $\theta_j$ (which counts $j$ rounding errors) is bounded by
\begin{equation}
\gamma_j = j\mu/(1-j\mu)\label{eq:gammaerroreq}
\end{equation}
and that this is largest when $k=\ell$, being $\gamma_{2\ell}$.  Notice that zero coefficients are not disturbed. This implicitly requires that $\ell$ not be so large that $2\ell\mu \ge 1$, which would happen only with impractically large degree polynomials.  See~\cite{Higham(1996)} or~\cite{corless2013graduate} for a proof of that fact and for more practice with the notation.

A similar result is true for blends. We have a proof, which relies on the positivity of the terms, and thus relies on $s$ being in the interval $[0,1]$, which will be published elsewhere.  We prove there that the floating-point evaluation of $H(s)$ in this interval by our adapted Horner algorithm will, if no overflow or underflow occurs, give the \emph{exact} value of a blend with different coefficients $p_j(1+t_j)$ and $q_j(1+s_j)$, where each $|t_j|$ is smaller than $\gamma_{3m+2n+4}$ and each $|s_j|$ is smaller than $\gamma_{2m+3n+4}$.  But, experimentally, more seems to be true: the backward stability seems to be true over quite a large region in the complex $s$-plane, not just on the interval $0 \le s \le 1$.

This means that the effects of rounding error can be modelled by the usual combination of backward error (guaranteed to be small) times conditioning.  We will see in the next section that blends are usually well-conditioned, and that balanced blends are the best.

\subsection{Conditioning and the Lebesgue function}
One common measure of the numerical behaviour of a polynomial expression in a given basis is the so-called \emph{Lebesgue} function of the basis: this is defined to be what you would get if absolute values are taken of each term multiplying a coefficient, and moreover all coefficients are also replaced by $1$.  More formally, if we expand $f(z)$ using the basis $\phi_k(z)$ for $0 \le k \le n$, so that
\begin{equation}\label{eq:generalf}
  f(z) = \sum_{k=0}^{n} a_k \phi_k(z)\>,
\end{equation}
then
\begin{equation}\label{eq:Lebesguefnex}
  | f(z) | = \left|\sum_{k=0}^{n} a_k \phi_k(z) \right| \le \max_{0 \le k \le n} |a_k| \sum_{k=0}^{n} |\phi_k(z)|
\end{equation}
and if we write $L(z) = \sum_{k=0}^n |\phi_k(z)|$ and call it the Lebesgue function for the basis $\phi_k(z)$ then the absolute value of $f(z)$ is bounded by the infinity norm of the vector of coefficients of $f(z)$ times the Lebesgue function.  This is simply the H\"older inequality applied to the expression for $f(z)$.

\begin{figure}
\centering     
\subfigure[Balanced case]{\label{fig:balancedL}\includegraphics[width=60mm]{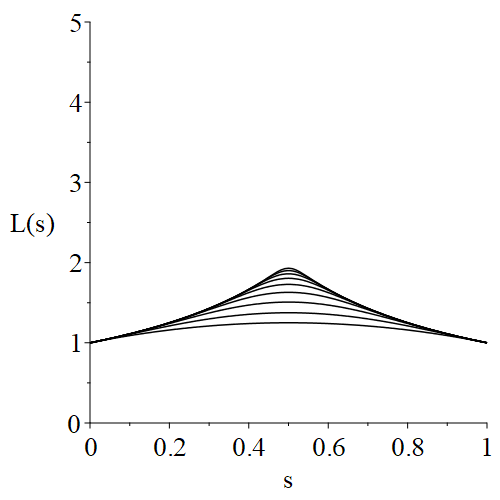}}
\subfigure[Unbalanced case]{\label{fig:unbalancedL}\includegraphics[width=60mm]{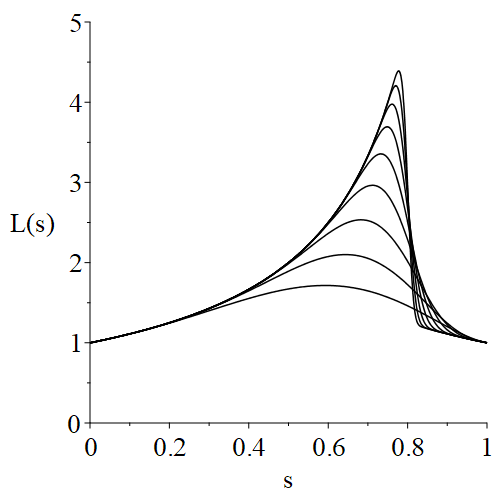}}
\caption{
Lebesgue functions for $(m,n) = 2^k$, $0 \le k \le 8$ (balanced case) and $(m,n) = (4\cdot2^{k},2^k)$, $0 \le k \le 8$ (unbalanced case). We see that in the balanced case, errors will be amplified by at most a factor of two; in the unbalanced case, it can be more, depending on the degree of unbalancing, but never more than the maximum of~$m$ and~$n$.
\label{fig:Lebesgue}}
\end{figure}

In our case we can see that on the interval $0 \le s \le 1$ all terms in the first series, with $p_j$, are positive anyway; in the second term, we may make everything positive by choosing $q_j = (-1)^j$.  Outside that interval, the absolute values are needed.
\begin{align}
    L_{m,n}(s) = &\sum _{j=0}^{m} \left\vert \sum _{k=0}^{m-j}{n+k\choose k}{s}^{k+j}
 \left( 1-s \right) ^{n+1}\right\vert   \nonumber\\
 +& \sum _{j=0}^{n}
 \left\vert\sum _{k=0}^{n-j}{m+k\choose k}{s}^{m+1}
 \left( 1-s \right) ^{k+j}\right\vert  \label{eq:Lebesgue}
\end{align}
Thus the Lebesgue function for our blend is (inside $0 \le s \le 1$) a blend itself, for a function with the same Taylor series at $s=0$ as $1/(1-s)$, and the same Taylor series at $s=1$ as $1/s = 1/(1 + (s-1))$. There is no function analytic everywhere with those properties, of course, but nonetheless these polynomials are useful.
Having a small size of $L$ is a guarantee of good numerical behaviour, if one implements things carefully.  Here, for the balanced case $m=n$, one can show that \emph{inside the interval} $1 \le L_{m,m}(s) \le 2$, no matter how large~$m$ is.  If~$m$ and~$n$ are large but not balanced, then we can have $L_{m,n}(s)$ as large as the maximum of~$m$ and~$n$.  See figure~\ref{fig:Lebesgue}.

Outside the interval $0 \le s \le 1$ the Lebesgue function grows extremely rapidly: not exponentially fast, but like a degree $m+n+1$ polynomial in $|s|$.  This essentially guarantees that blends are typically useful numerically only between the two endpoints and in a small region in the complex plane surrounding that line segment; that is, where $L(s)$ remains of moderate size.  By refining this argument somewhat, we may do better for certain polynomials by taking better account of the polynomial coefficients through the theory of conditioning, see~\cite{corless2013graduate}.  We do not pursue this further here.

\section{Integration of a blend}
We will now see that the \emph{definite} integral of a blend over the entire interval will allow us to construct a new blend whose value at any point is the \emph{indefinite} integral of the original blend up to that point, $F(x) = \int_{s=0}^x H_{m,n}(s)\,ds$.

Direct integration over the entire interval $0 \le s \le 1$ and use of the formula
\[
\int_{s=0}^1 s^{a}(1-s)^b\,ds = {\frac {a!\,b!}{ \left( a+b+1 \right) !}}
\]
gets us a formula for $F(1)$ involving the symbolic sum
\begin{equation}
  \sum _{k=0}^{m-j}{\frac {{n+k\choose k} \left( j+k \right) !\, \left(
n+1 \right) !}{ \left( n+2+j+k \right) !}}
\end{equation}
and a similar one interchanging~$m$ and~$n$.
\Maple\  can evaluate both those sums:
\begin{lstlisting}
sm := sum( binomial(n+k,k)*(n+1)!*(k+j)!/(j+k+n+2)!, k=0..m-j ):
simplify( sm );
\end{lstlisting}
yields the right-hand side of the equation below:
\begin{equation}\label{eq:closedsum1}
  \sum _{k=0}^{m-j}{\frac {{n+k\choose k} \left( j+k \right) !\, \left(
n+1 \right) !}{ \left( n+2+j+k \right) !}} = {\frac { \left( n+m-j+1 \right) !\, \left( 1+m \right) !}{ \left( j+1
 \right)  \left( n+2+m \right) !\, \left( m-j \right) !}}
\>.
\end{equation}
Similarly we find the other sum, and finally we get
\begin{align}
\int_{s=0}^1 H_{m,n}(s)\,ds = &
{\frac { \left( m+1 \right) !}{ \left( m+n+2 \right) !}\sum _{j=0}^{m}
{\frac { \left( n+m-j+1 \right) !}{ \left( j+1 \right)
 \left( m-j \right) !}}}\,p_{{j}}\nonumber\\
 &+{\frac { \left( n+1 \right) !}{ \left( m+n+2
 \right) !}\sum _{j=0}^{n}{\frac {  \left( n+m-j
+1 \right) !}{ \left( j+1 \right)  \left( n-j \right) !}}}\,\left( -1 \right) ^{j}q_{{j}}
\>.
\label{eq:sumasintegral}
\end{align}
The numbers showing up in this formula turn out to be smaller for the higher-order Taylor coefficients, as one would expect.  We emphasize that the above formula gives (in exact arithmetic) the \emph{exact} integral of the blend over the whole interval.  If the blend is approximating a function $f(s)$, then integrating equation~\eqref{eq:errorblend} gives us
\begin{equation}\label{eq:integrationerror}
  \int_{s=0}^1 f(s)\,ds - F(1) = (-1)^{n+1}\frac{(m+1)!(n+1)!}{(m+n+3)!} \frac{f^{(m+n+2)}(c)}{(m+n+2)!}
\end{equation}
where, using the Mean Value Theorem for integrals and the fact that $s^{m+1}(1-s)^{n+1}$ is of one sign on the interval, we replace the evaluation of the derivative at one unknown point $\theta$ with another unknown point $c$ on the interval.

Once we have the value $F(1)$, we can construct a new blend from the old one as follows.  First, we put a value of $0$ for the new $F(0)$ at the left end (in a string of blends, we would accumulate integrals; for now, we are just integrating from the left end).  Then we adjust all the Taylor coefficients at the left: the old $f(0)/0!$ becomes the new $F'(0)/1!$, the old $f'(0)/1!$ becomes the new $F''(0)/2!$ and so we have to divide the old $p_1$ by $2$; the old $f''(0)/2!$ becomes the new $F'''(0)/3!$ so we have to divide by $3$, and so on until the old $f^{(m)}(0)/m!$ becomes the new $F^{(m+1)}(0)/(m+1)!$; the new blend will have $m+2$ Taylor coefficients on the left (indexing starts at $0$).

Now we make $F(1) = $ the integral given above. We then shift all the old $q_j = f^{(j)}(1)/j!$ into the new $F^{(j+1)}(1)/(j+1)!$ for $j=0$, $\ldots$,~$n$.

We now have a type $(m+1,n+1)$ blend $H_{m+1,n+1}(s)$.  Its Taylor coefficients on the left are the same as the Taylor coefficients of $F(x) = \int_{s=0}^x H_{m,n}(s)\,ds$ as a function of $x$.  Its Taylor coefficients on the right are also the same as those of $F(x)$ at $x=1$.  Thus we have a blend for the integral.  Its grade is $m+1+n+1+1$ which is $m+n+3$, not $m+n+2$.  However, in exact arithmetic, the result is actually of degree at most $m+n+2$, because the value is the exact integral of a polynomial, and thus we see that the blend we have is actually using more information than it needs.  We \emph{could} throw one of the highest derivatives away (it's natural to do so at the right end) but there is no real need
unless we expect to do this process repeatedly to a single blend.
%

To use this formula on integration from $z=a$ to $z=b$ one must incorporate the change of variable from $z$ to $s = (z-a)/(b-a)$.  Putting $h=(b-a)$ then we must (as always) scale the Taylor coefficients $p_j$ and $q_j$ by multiplying each by $h^j$, and then finally the integral is just
\begin{equation}\label{eq:scaledintegral}
  \int_{z=a}^b H_{m,n}\left( \frac{z-a}{b-a} \right)\,dz = h\int_{s=0}^1 H_{m,n}(s)\,ds \>.
\end{equation}

If we have more than one blend lined up in a row, which we call a ``string of blends'' (this is quite natural, as can be seen from the fact that blends are joined at what are termed ``knots'' in the spline and piecewise polynomial literature), then this formula can be used to generate composite quadrature rules.  The case $m=n=0$ just gives the trapezoidal rule, which is right because the blend is just a straight line; if instead $m=n=1$ then we get what is called the ``corrected trapezoidal rule''
\begin{equation}\label{eq:scaledintegral}
  \int_{z=a}^b H_{1,1}\left( \frac{z-a}{b-a} \right)\,dz = \frac{h}{2}\left( f(a) + f(b) \right) + \frac{h^2}{12}\left( f'(b) - f'(a) \right) \>.
\end{equation}
A $(4,4)$ blend gives the rule
\begin{align}\label{eq:fourblendint}
  \int_{s=0}^1 H_{4,4}(s)\,ds = & {\frac {p_{{0}}}{2}}+{\frac {p_{{1}}}{9}}+{\frac {p_{{2}}}{36}}+{
\frac {p_{{3}}}{168}}+{\frac {p_{{4}}}{1260}}\nonumber \\
   & +{\frac {q_{{0}}}{2}}-{
\frac {q_{{1}}}{9}}+{\frac {q_{{2}}}{36}}-{\frac {q_{{3}}}{168}}+{
\frac {q_{{4}}}{1260}}
\end{align}
To get a valid rule on an interval of width $h$, one needs powers of $h$ in the Taylor series.  We see that this balanced blend gives coefficients that will telescope at odd orders for composite rules on equally-spaced intervals. [This is well-known.]  See also~\cite{talvila2016higher} for optimal formulas of this balanced type.
\section{Horner Form}
If we look at equation~\eqref{eq:TPHI} with a programmer's eye, we see a lot of room for economization. First, the sums are polynomials in~$s$ and in $1-s$.  Because $0 \le s\le 1$, both of these terms are positive, so we do not want to expand powers of $(1-s)$, for instance; introducing subtraction means potentially revealing rounding errors made earlier.  But as a first step we may put the sums in Horner form.  We remind you that the \emph{Horner form} of a polynomial $f(x) = f_0 + f_1 x + f_2 x^2 + f_3 x^3$ is a rewriting so that no powers occur, only multiplication: $f(x) = f_0 + x(f_1 + x(f_2 + x f_3))$.  The form can be programmed in a simple loop:
\begin{lstlisting}
p := f[n];
for j from n-1 by -1 to 0 do
  p := f[j] + x*p;
end do;
\end{lstlisting}
Here we have a double sum, and in each sum we may write in Horner form; that is, where the loop above has a simple \lstinline!f[j]! we would have an inner Horner loop to compute it.

But the inner sum is simply $\sum_{k=0}^{m-j} \binom{n+k}{k} s^k$ once the $s^j(1-s)^{n+1}$ is factored out of it.  These inner sums should be precomputed by the simple recurrence (adding the next term to the previous sum), outside of the innermost loop, so that the cost is proportional to either~$n$ or~$m$, and not their product.

The numbers $\binom{m+k}{k}$ and $\binom{n+k}{k}$ occur frequently, and perhaps they should be precomputed.  Except that they, too, can be split in a Horner-like fashion, because for $k\ge 1$
\[
s^k\binom{m+k}{k} = s\frac{m+k}{k} \cdot s^{k-1}\binom{m+k-1}{k-1}\>.
\]
While this is actually more expensive than precomputing the numbers, by keeping $s$ involved, the loop keeps the size of the numbers occuring in the formula small (remember $0 \le s \le 1$), and this contributes to numerical stability.  This is best seen by example.  In the $m=n=3$ case, one of the terms is
\[
1 + 4s + 10s^2 + 20s^3
\]
which rewritten in Horner form is just $1+s(4+s(10+s\cdot 20))$.  But if we factor out the binomial coefficient factors using the rule above, it becomes
\[
1+4\,  s\left( 1+\frac52\, \left( 1+2\,s \right) s \right)\>.
\]
It might be better to keep only integers in the rewritten form; we do not know how to do that in general, although it is simple enough for this example.

A final and important efficiency is to realize that the sum for the left-hand terms and the sum for the right-hand terms is invariant under a symmetry: exchange~$m$ and~$n$, exchange~$s$ and $1-s$, and account for sign changes in the second sum by absorbing them into the $q_j$, and the sums can be executed by the same program. This leads to later \emph{programmer} efficiency as well, if one thinks of a further improvement to the code: then it only has to happen in one place. [This actually happened here.]  We give the algorithm for this half-sum in Algorithm~\ref{alg:Horner}.  To compute the blend, this algorithm is called once with $m$, $n$, and $s$ and the coefficient vector $p_j$, and once with $n$, $m$ (note the reverse order), $1-s$, and the coefficient vector $(-1)^jq_j$, and the results are added.

The goal is to make the innermost loop as efficient as is reasonably possible.  We expect that these blends will be evaluated with hundreds of points (routinely) and on occasion with tens of thousands of points (for a tensor product grid of a bivariate function, for instance).  In \Maple, we want to be able to use \lstinline!evalhf! or even the compiler.  This provides significant speedup.  

\begin{algorithm}[h!]
\caption{Horner's algorithm adapted for one of the two sums of the blend.}
\label{alg:Horner}
\begin{algorithmic}[1]
\Procedure{HSF}{$m$, $n$, $\sigma$, $w$ }
  \State $a_0 \gets 1$
  \For{ $k \gets 1\ldots m$}
    \State $a_k \gets (n+k)\sigma a_{k-1}/k$  
  \EndFor
  \For{ $k \gets 1$ to  $m$}
    \State $a_k \gets a_{k-1} + a_k$  
  \EndFor
  \State $u \gets 0$
  \For{ $j \gets m$ by $-1$ to $0$}
    \State $u \gets a_{m-j} w_j + \sigma u$ 
  \EndFor
  \State $c \gets 1$
  \For{ $j \gets 1$ to $n+1$}
    \State $c \gets (1-\sigma)c$  
  \EndFor
  \State $e \gets cu$ 
  \State \textbf{return} $e$
\EndProcedure
\end{algorithmic}
\end{algorithm}

\subsection{``Automatic'' differentiation}
The Horner loop above can be rewritten to provide not only the value of $p(x)$ but also of $p'(x)$, the derivative with respect to $x$.  This is also called program differentiation.  \Maple's \lstinline!D! operator can differentiate simple programs such as that. Supposing we define
\begin{lstlisting}
Horner := proc(x, f, n)
  local i, p;
  p := f[n];
  for i from n-1 by -1 to 0 do
    p := f[i] + x*p;
  end do;
  return p;
end proc:
\end{lstlisting}
Then the command \lstinline!D[1](Horner)! produces the following:
\begin{lstlisting}
proc(x, f, n)
    local i, p, px;
    px := 0;
    p := f[n];
    for i from n - 1 by -1 to 0 do
        px := px*x + p;
        p := f[i] + x*p;
    end do;
    px;
end proc
\end{lstlisting}
This procedure returns just the derivative, not the derivative and the polynomial value.  If one wishes that, one may use instead \lstinline!codegen[GRADIENT]!, with the syntax
\begin{lstlisting}
codegen[GRADIENT](Horner, [x], function_value = true)
\end{lstlisting}
This command generates the following code.
\begin{lstlisting}
proc(x, f, n)
    local dp, i, p;
    dp := 0;
    p := f[n];
    for i from n - 1 by -1 to 0 do
        dp := dp*x + p;
        p := p*x + f[i];
    end do;
    return p, dp;
end proc
\end{lstlisting}
Procedures for evaluating higher-order derivatives may be computed in a similar way.

For our purposes, though, it is better to allow an \emph{arbitrary} number \lstinline!nder! of derivatives. This means not adding one or more statements to the Horner loop, but rather writing a loop to evaluate all the derivatives.  Here is this idea applied to the Horner program above.
\begin{lstlisting}
Horner := proc(x, f, n, nder)
  local i, ell, p;
  p := Array(0..nder,0);
  p[0] := f[n];
  for i from n-1 by -1 to 0 do
    for ell from nder by -1 to 1 do
      p[ell] := p[ell]*x + ell*p[ell-1];
    end do;
    p[0] := f[i] + x*p[0];
  end do;
  return p;
end proc:
\end{lstlisting}
Calling this with \emph{symbolic} $x$ and $f$ and numeric~$n$ and number of derivatives desired, gets something like (for $n=3$)
\[
p(x) = f_{{0}}+x \left( f_{{1}}+x \left( xf_{{3}}+f_{{2}} \right)  \right)
\]
which looks familiar, but has the expression
\[
p'(x) = \left( 2\,xf_{{3}}+f_{{2}} \right) x+f_{{1}}+x \left( xf_{{3}}+f_{{2}
} \right)
\]
which looks strange.  But it's correct---just a rewriting of the normal derivative of a cubic polynomial.  But the strength of this technique is not for symbolic use, but rather for numeric use.  When calling the modified program with a numeric $x$ then the loop just performs (reasonably efficient) numerical computation; this program can be translated into other languages, as well.

For the code for our blends, we simply wrote all the loops ourselves as above.  We have not yet tried to translate the resulting code (which is more complicated than the simple Horner loop above) into any other languages.
\subsection{User interface considerations}
There is similar code in the \lstinline!Interpolation! package and in the \lstinline!CurveFitting! package, namely \lstinline!Spline! and \lstinline!ArrayInterpolation!.  The interface to this code should not be too much different to those.  Consideration of the various possible kinds of inputs demonstrates that a front-end that dispatches to the most appropriate routine would be helpful; if the input~$s$ is a symbol, then there is no point in calling \lstinline!evalhf!, for instance.  If the input is an \lstinline!Array! of complex floating-point numbers, then depending on Digits it might indeed be appropriate to try the hardware float routine.

For that reason we chose a \lstinline!module! with \lstinline!ModuleApply! as being most convenient; this would allow the user to be relatively carefree.  We also allowed the module to export the basic `fast' routines so that if the user wanted to look after the headaches of working storage of hardware floating point datatypes then the user could use blends in their own code without a significant performance penalty.

Another issue is a good name.  We chose \lstinline!HornerTwoPointHermiteInterpolation! before we thought of the name \emph{blend}.  At the moment we have kept the old name and use \lstinline!macro(Blend=HornerTwoPointHermiteInterpolation)! for short.

The minimum information that the routine needs is $z$, $a$, $b$, and the Arrays $p$ and $q$ of Taylor coefficients. If the user does not request a number of derivatives, it can be safely assumed that only $H(z)$ is wanted.  The grade $(m,n)$ of the blend can be deduced from the input Arrays $p$ and $q$.  As a convenience to the user we allow the ability to specify~$m$ or~$n$ even if the input Arrays are larger.

The types of data input can vary considerably.  We allow rationals, exact numerics, software floats, hardware floats, and complex versions of all of those.  We do not provide for finite fields (the binomial coefficients would in some cases then possibly be zero---and we don't even know if formula~\eqref{eq:TPHI} is even true over finite fields---in that case) or for matrix values although for that latter case the concept is well-defined.

The data type of the output can vary, as well: when there is an Array of inputs, and only function values are wanted and no derivatives, the user would surely expect an Array of outputs of the same dimension.  If derivatives are wanted, though, then there will be a higher-dimensional Array output; sometimes the special case of an index~$0$ for such a higher-dimensional output would fit the user's expectations so we allow an option to specify such.  The default is just to be sensible: scalar in, scalar out; Vector in, Vector out.

Currently several operations take place outside the code, in ``main \Maple''.  This includes series manipulations and the construction of the companion matrix pair.  Construction of the integrated blend is also currently left in the user's hands.

At this moment we do not know just how this code would be used, or who would be interested (aside from people interested in high-order methods for solving differential equations numerically). The idea is surprisingly flexible: by reversing Taylor series at each end, it is easy to make blends of inverse functions, for instance.  The usage will affect how convenient or inconvenient the user interface is.  So we tried to make the interface as simple as possible. Doubtless improvements will occur to us after it's out in the wild.
%
\section{Testing and Timing\label{sec:timing}\label{sec:testing}}
\begin{figure}
  \centering
  \includegraphics[width=6cm]{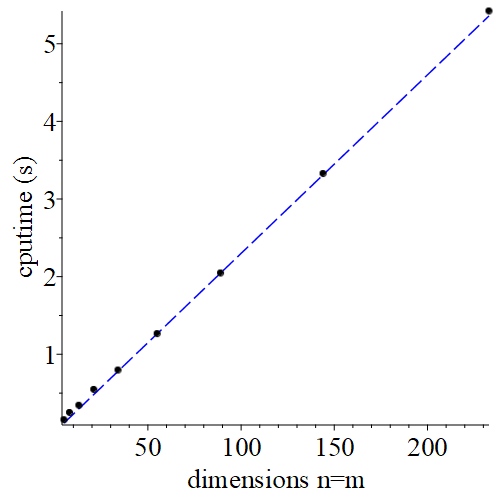}
  \caption{CPU time in seconds on an i5-7300U 2.6Ghz  Microsoft Surface Pro running \Maple\  2020 for balanced blends $H_{m,m}(s)$ of varying degrees.  The blends and their first three derivatives were evaluated on $2021$ equally spaced points on the interval $0 \le s \le 1$ including the endpoints. The fitted curve is $0.0023m$ showing linear growth of computing time, as should have been expected. }\label{fig:cputimes}
\end{figure}

In figure~\ref{fig:cputimes} we see the results of a simple test with random Taylor coefficients drawn from the interval $-1 \le x \le 1$.  We first used the \Maple\ \lstinline@rand@ function to generate coefficients for the maximum~$m$ and~$n$. Subsequent calls to \Blend\ used subsets of those data.  The blends were evaluated in $15$ Digit precision at $2021$ points equally-spaced on $0 \le s \le 1$ including the endpoints. The code was asked to compute derivatives up to order~$3$.  That is, four quantities were computed at each point: $H_{m,m}(s)$, $H'_{m,m}(s)$, $H''_{m,m}(s)$, and $H'''_{m,m}(s)$.  The computing time was modest and showed linear growth, with a fit of $0.023m$ to its data (in seconds). Thus the computing time seems, as expected, linear in the degree of the balanced blend.  We ran a further test with the same coefficients but this time without asking for derivatives; the cost (not shown) was a factor $4.2$ less.  In both cases the main call was used, so these times include the times for preparation and dispatch to evalhf.

\begin{figure}
\centering     
\subfigure[$f(s)-H_{8,8}(s)$]{\label{fig:cos88a}\includegraphics[width=60mm]{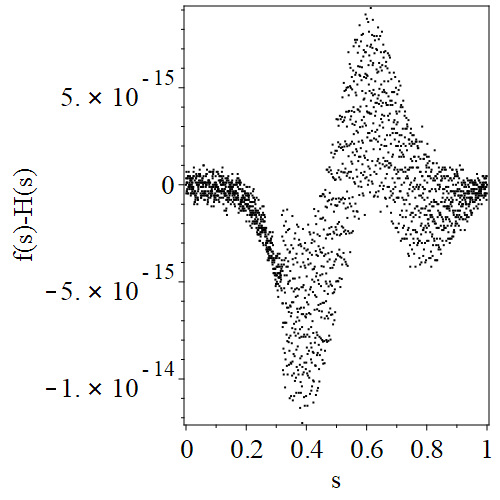}}
\subfigure[$(f'(s)-H_{8,8}'(s))/\pi$]{\label{fig:cos88b}\includegraphics[width=60mm]{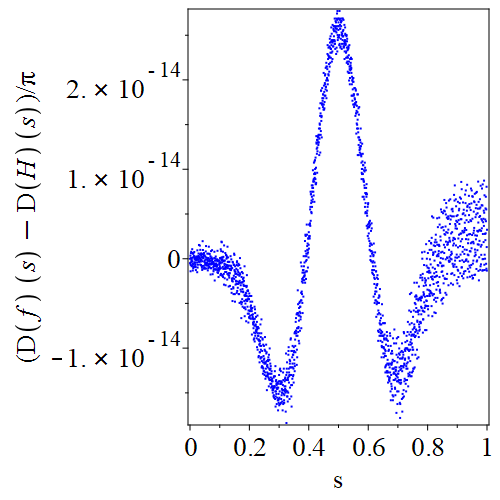}}
\newline
\subfigure[$(f''(s)-H_{8,8}''(s))/\pi^2$]{\label{fig:cos88c}\includegraphics[width=60mm]{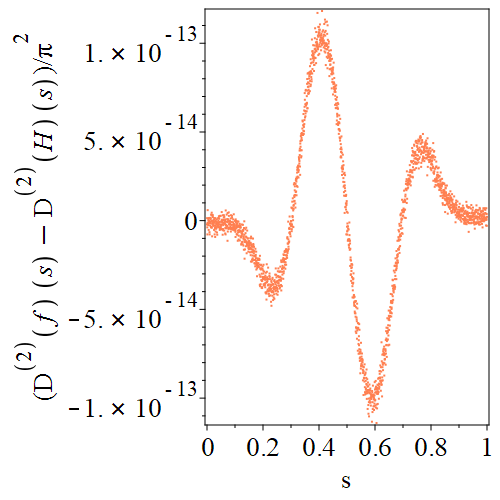}}
\subfigure[$(f'''(s)-H_{8,8}'''(s))/\pi^3$]{\label{fig:cos88d}\includegraphics[width=60mm]{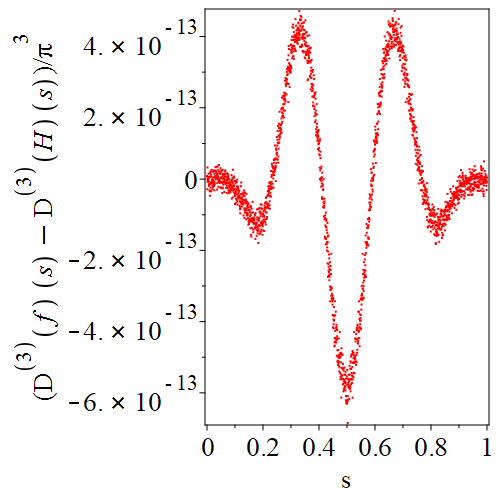}}
\caption{The error in an $(8,8)$ blend for $f(s) = \cos\pi\,s$.  This grade of blend produces an approximation that is nearly accurate to full double precision; the truncation error, proportional to $s^9(1-s)^9$, is beginning to be obscured by rounding errors.  Recomputing these errors at higher precision gives smoother curves of about the same size. As usual with approximation methods, the accuracy degrades as the derivative order increases. \label{fig:cos88}}
\end{figure}

For testing stability and accuracy, we first looked at very smooth functions.  In figure~\ref{fig:cos88} we see error curves at $15$ Digits for $(8,8)$ blends for $f(s) = \cos\pi s$ and its derivatives. This shows the effects, scaled with the appropriate power of $\pi$, of taking the derivative.
This function has known Taylor series at each end (indeed the coefficients are just the negatives of each other): $\cos\pi s = 1 - (\pi s)^2/2! + (\pi s)^4/4! - \cdots$ and at the other end $\cos \pi s = $
\[
 -1+{\frac {{\pi}^{2}}{2}} \left( s-1 \right) ^{2}-{\frac {{\pi}^{4}}{
24}} \left( s-1 \right) ^{4}+{\frac {{\pi}^{6}}{720}} \left( s-1
 \right) ^{6}-{\frac {{\pi}^{8}}{40320}} \left( s-1 \right) ^{8}+O
 \left(  \left( s-1 \right) ^{10} \right) \>.
\]
The $(9,9)$ blends are better---and use essentially the same information because the Taylor coefficients of degree $9$ at either end are zero---but these curves are informative about the numerical stability and efficiency of these blends.

\begin{figure}
  \centering
  \includegraphics[width=6cm]{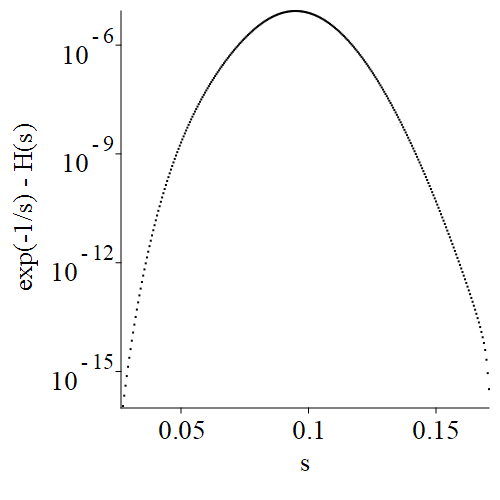}
  \caption{The difference between $\exp(-1/s)$ and its $(900,100)$ blend near its maximum point. The maximum difference is about $10^{-5}$. Computing this grade $1001$ blend and its derivative at $2021$ points (once the series coefficients at $s=1$ were computed; the series coefficients at $s=0$ are all $0$) took under $13$ seconds on an i5-7300U 2.6Ghz Microsoft Surface Pro. Note that $\binom{1000}{100}\approx 6.4\cdot 10^{139}$ but no numerical artifacts are seen: the blend is smooth all across $0 \le s \le 1$.}\label{fig:smooth900100}
\end{figure}

We then chose a harder example.  In figure~\ref{fig:smooth900100} we find the results of a ``stress test'', namely a blend for the function $f(s) = \exp(-1/s)$.  This has all its right-sided derivatives at $s=0^+$ being zero, but the function is not analytic there, and indeed has an essential singularity there.  At the other end, $s=1$, we use Maple's symbolic-order differentiation capability~\cite{benghorbal2002n} \lstinline!diff(exp(-1/x),x$k)! to find that, for $k\ge 1$,
\begin{equation}\label{eq:symbolder}
  \left.\frac{d^{(k)}f}{dx^k}\right\vert_{x=1} = (-1)^{k+1}e^{-1/2} \mathrm{WhittakerM}\left(k,\frac12,1\right) = (-1)^{k+1}e^{-1}F\left(\left.{1-k \atop 2}\right\vert 1\right)\>.
\end{equation}
Here $F$ represents \lstinline@hypergeom( [1-k], [2], 1)@.
For $k=0$ one uses the same formula but adds $1$.  Maple knows how to evaluate these; they are rational multiples of $\exp(-1)$.  It is amusing to note that apart from sign, the first $5$ are just $\exp(-1)/k!$, but the degree $5$ term is $-19\exp(-1)/5!$: only computing to degree $4$ could have led to a false experimental conclusion! This formula allows us to compute as many series coefficients at $s=1$ as we could wish.  We take $n=900$, and $m=100$, giving a grade $1001$ blend.  Indeed only the $q$ part of formula~\eqref{eq:TPHI} is present, so the blend is actually degree $1001$ not just grade $1001$.  The largest binomial coefficient appearing is $\binom{1000}{100}$ which is about $6.4\cdot 10^{139}$ which suggests that numerical difficulties are to be expected.  None, however, appear. The blend is entirely smooth, and the difference between the blend and $f(s)$ is no more than $10^{-5}$ at its greatest.  One expects that the blends will converge as $(m,n)$ go to infinity, for any fixed ratio of~$m$ and~$n$.  Here because the ratio was $9/10$ we find the maximum error occurring near $s=0.1$ (about $s=0.095$).

It is natural to compare with the pure Taylor series at $s=1$, both of degree $900$ and of $1001$.  The errors at $s=0$ are, naturally, far larger, because that series diverges there. The degree $900$ polynomial has error $-0.0558$, while the degree $1001$ polynomial has error $0.0576$.  The blend wins very handily.

\begin{figure}
  \centering
  \includegraphics[width=8cm]{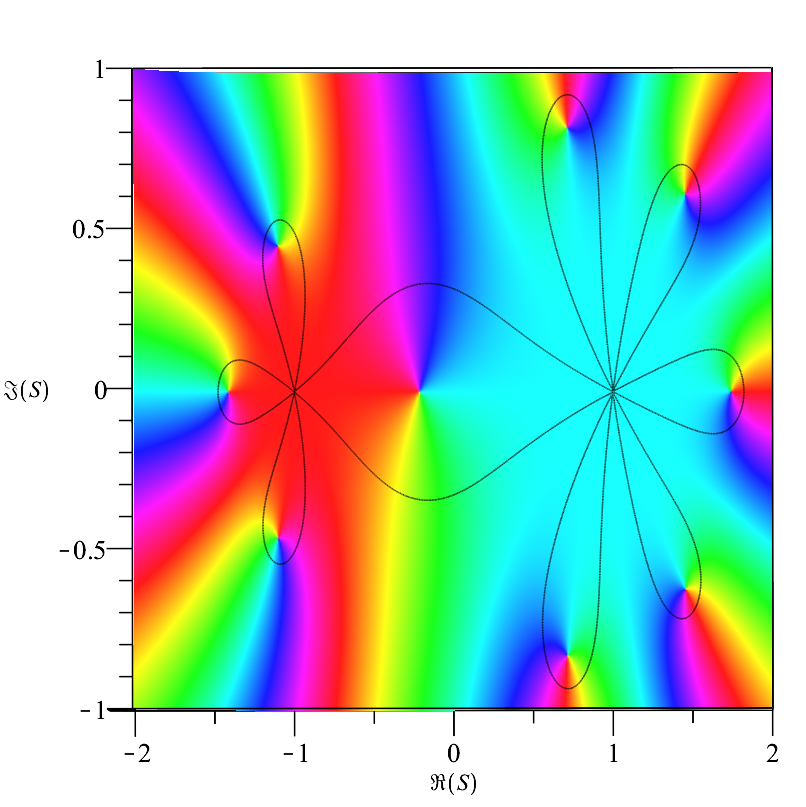}
  \caption{Phase plot (see~\cite{wegert2012visual}) of the blend $H_{3,5}(s)$ of a step function over a region of the complex plane surrounding the unit interval.}\label{fig:Flag35}
\end{figure}

We now give another stress test, this one (finally) showing some numerical failure (overflow and underflow, resulting in NaNs, or floating-point (Not A Number)s).  We blend the step function $f(s)=-1$ at $s=0$ with all derivatives zero and $f(s)=1$ at $s=1$ with all derivatives zero.  Depending on the ratio of~$m$ and~$n$, the step will be located somewhere between; near $s = (m+1)/(m+n+2)$ in fact.  The Lebesgue function is maximal at that point, with value $(\max(m,n)+1)/(\min(m,n)+1)$.  In figure~\ref{fig:Flag35} we see a phase plot in the complex plane of a modest $(3,5)$ blend for this function; this was symbolically computed and plotted with no difficulty using the code below.  

\begin{lstlisting}
Digits := 15: 
(m,n) := (3,5): 
p := Array(0..m,[-1,0$m]): 
q := Array(0..n,[1,0$n]): 
H := Blend( S, -1, 1, p, q ):  
plots:-complexplot3d(H, S = -2-I .. 2+I, style=surfacecontour, 
                        contours=[1], orientation=[-90,0], 
                        lightmodel=none, size=[0.75, 0.75], 
                        grid=[400,400] );

\end{lstlisting}

\begin{figure}
  \centering
  \includegraphics[width=6cm]{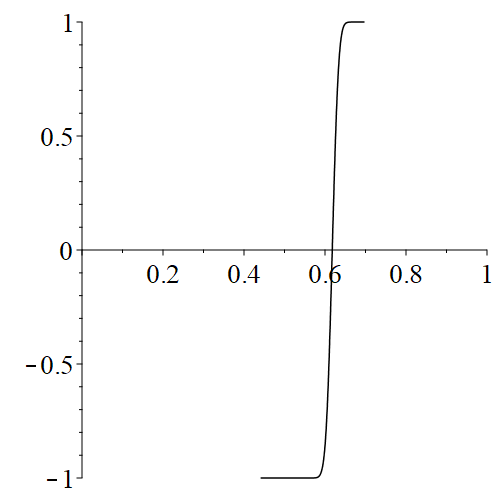}
  \caption{The $(987,610)$ blend of the step function $-1$ at $0$ and $1$ at $1$.  We finally see evidence of numerical difficulty: $\binom{987+610}{610}\approx 3.5\cdot 10^{459}$ and this causes both \emph{overflow} and \emph{underflow} resulting in NaNs, which are not plotted.  Every numerical value that is plotted is correct to $13$ digits, however: the only numerical failure is overflow.  This blend has grade $1598$ and computing it and its derivative at $2021$ values (many of which resulted in NaNs) took $7.7$ seconds on an i5-7300U 2.6Ghz Microsoft Surface Pro.   }\label{fig:step1615}
\end{figure}

For higher grades, we must use the numerical code of this paper.  By taking very high grades, we stress the floating-point capacity of the code.
For $(m,n) = (987,610)$ (these are Fibonacci numbers, by the way) we have the largest binomial coefficient about $3.5\cdot 10^{459}$ which must overflow in IEEE double precision, which is used by \lstinline@evalhf@.  The corresponding powers of~$s$ and $1-s$ must underflow. In spite of that, the blend correctly computes (taking $7.7$s CPU time) the step portion of the figure: overflow and underflow causing NaNs only happen in the flat portions of the blend.  See figure~\ref{fig:step1615}.  Computing instead in $30$ Digits (which takes about $70$ seconds on the same machine) does not suffer from overflow or underflow because software floats in \Maple\ have a greater range.  At this precision, \Maple\ computes the complete figure (not shown).  Moreover, comparing the numerical values computed at $15$ Digits to the values computed at $30$ Digits, we find that the largest difference is smaller than $7\cdot 10^{-14}$.  Working at $15$ Digits, the blend was able to correctly compute the interesting part of the curve, even though overflow/underflow prevented it from computing the flat parts. The derivative was computed in $15$ Digits correct to $10^{-11}$ in the same region the function was computed correctly.

Note that $\phi\cdot 2020 \approx 1248.4$ where $\phi = (\sqrt{5}-1)/2 \approx 0.6180$ is the golden ratio.  If the output of the call to \Blend\ was stored in the \lstinline@Array(0..2020,0..1)@ $y$, then the value of $y[1248,0] = -0.0074$ while $y[1249,0] = 0.0250$, indicating that the location of the step is indeed determined by the ratio of $m+1$ to $m+n+2$, being Fibonacci numbers.

We conclude that blends of degrees high enough to produce binomial coefficients $\binom{m+n}{n}$ that overflow will cause numerical difficulty.  What seems surprising is that this is the only case where we have seen numerical difficulty.  We have also looked at cases where the function has nearby complex poles and so the series cannot converge, and while the blends do have unexpected features in the regions between the two points of expansion of the series, in all cases they behaved smoothly.  Even when we tried series for functions whose Taylor series are known to be ill-conditioned (such as $\exp(-50s)$ blending with $\exp(150(s-1)/\pi$) the resulting behaviour was explainable.

\section{Future Work}
The idea of blending two Taylor series is quite old, and people have tried to do it in several different ways.  The Hermite interpolation idea is one of the oldest, but we think that not enough attention has been paid to it.  There are other ways in the literature. For example, there is the \emph{very} similar work in~\cite{hummel1949generalization}, which makes a kind of blend with a variable upper limit and uses that to construct \emph{rational} approximations.

The next step of course is to combine different blends into what we call a string of blends, joined at ``knots'' where the same local Taylor series are reused.  This is a kind of \emph{piecewise polynomial}, similar to splines which are another kind of piecewise polynomial.  Of course having a single blend use more than two truncated Taylor series is just Hermite interpolation. There are a great many other similar ideas in the literature.

It is not clear if what we are calling a ``blend'' will be sufficiently useful to catch on widely; the existing body of numerical software involving piecewise polynomials is quite substantial, and it is not clear that a blend is any better than what is being used already.  However, there are some niche situations, such as numerical solution of ODE by high-order methods, where it is natural (and already being used in specialized software).  There may also be useful pedagogical reasons to talk about blends (whimsically, we called some of the unusual quadrature formulas ``Anti-Cheating Quadrature rules''; this may be of interest for student assessments!). The idea of a blend does reinforce the ideas of convergence and approximation.  Trying to blend two series that have complex poles near to each expansion point produces some very informative results!  The answer to ``Will it blend?'' is, in that case, ``no''.  The antics of the blends as they try to converge (when they \emph{can't}) is quite entertaining.


We would like to extend this code to vector and matrix blends.  We would also like to blend Laurent and Puiseux series (Laurent series seem, in fact, very simple: just do a blend of the Taylor series for $(z-a)^\alpha(z-b)^\beta f(z)$---but we haven't tried this yet).  Creating an environment where one can add, subtract, multiply, and apply other operations to blends and produce new blends, might be of interest, in a way similar to Chebfun (\href{www.chebfun.org}{www.chebfun.org}).  It is in such an environment where the companion pair and the integration of blends would fit most naturally.

Extending this work to the multivariable case (aside from the use of tensor product grids) may also be of interest.  However, the residue argument breaks down for finding formulas; other approaches will have to be used.  There is a significant literature on low-degree multivariate Hermite interpolation, and in the context of finite element methods there is interest in higher-degree methods (and considerable published work) as well.
%
%
%
\bibliographystyle{splncs04}
\bibliography{blendsandsplines}
\end{document}